# A New Enhanced Version of The Ensemble Interpretation

Raed M. Shaiia

Department of Physics, Damascus University, Damascus, Syria

**Abstract**   The different interpretations of quantum mechanics yield the same experimental results, which may give the impression that the question of what interpretation is the true one, is a philosophical question, not a scientific one. But in this paper, we will see that we can actually prove one interpretation, in particular, a version of the ensemble interpretation, as the natural interpretation of quantum mechanics. Furthermore, we will prove the axioms of quantum mechanics, without the need of anything beyond probability theory.

## 1. Introduction

Throughout this paper, Dirac's notation will be used.

In quantum mechanics, for example [1] in the experiment of measuring the component of the spin of an electron on the Z-axis, and using as a basis the eigenvectors of $\hat{\sigma}_3$, the state vector of the electron before the measurement is:

$$|\psi\rangle = \alpha|u\rangle + \beta|d\rangle \qquad (1)$$

However, after the measurement, the state vector of the electron will be either $|u\rangle$ or $|d\rangle$, and we say that the state vector of the electron has collapsed [1]. The main problem of a measurement theory is to establish at what point of time this collapse takes place [2].

Some physicists interpret this to mean that the state vector is collapsed when the experimental result is registered by an apparatus. But the composite system that is constituted from such an apparatus and the electron has to be able to be described by a state vector. The question then arises when will that state vector be collapsed? [1] [2].

On the other hand, if we consider the experiment of tossing a coin for one time, then the sample space of this simple experiment is[3]:

$$\Omega = \{H, T\} \qquad (2)$$

Of course, as it is known, this does not mean that the coin has all of these possibilities at once, it is merely a statement about the possible outcomes of the experiment. And after doing the experiment, we will get just one of these two results and not both. That means that one of the elementary events only will happen: either $\{H\}$ or $\{T\}$, but not both at once[3].

What if the state vectors were nothing but another representation of events in the sense of the usual probability theory? What if the state vector before measurement is just the representation of a sample space, and the result we get after measurement is just an ordinary elementary event, in a similar manner to the coin example, and in this sense the measurement is an experiment in the sense of the word used in probability theory? If we could reformulate probability theory in such a way that allows the representation of ordinary events by vectors, without violating Bell's theorem, then this will lead to an entirely different understanding of the underlying mathematics of quantum mechanics, and hence to quantum mechanics itself. And this is the aim of this paper.

## 2. An alternative method to formulate probability theory

In this section we focus on the reformulation of probability theory. Next, we use this formulation to reformulate quantum mechanics. All systems that we will study at first are classical, until it is otherwise stated. We reformulate probability theory in a similar language to the one used in quantum mechanics. Later on, we show that this formulation reduces the number of postulates used in quantum mechanics.

First we will start by considering finite sample spaces.
Here it will be presented an outline of the method to be used in this formulation:

Having an experiment with a finite sample space $\Omega$, There is always a finite dimensional Hilbert space $H$ with a dimension equal to the number of the elementary events of the experiment.

Then, we can represent each event by a vector in $H$ using the following method:

I- the square of the norm of a vector representing an event is equal to the probability of the event.

II- Given an orthonormal basis of $H$, we represent each elementary event by a vector parallel to one of these basis vectors, such that no different elementary events are represented by parallel vectors, and the square of the norm of the representing vector is equal to the probability of the elementary event.

III- Then every event is represented by the vector sum of the elementary events that constitute it.

From (I) we see that the vector $|\phi\rangle$ representing the impossible event must be the zero vector because:

$$\langle\phi|\phi\rangle = \|\phi\|^2 = p(\phi) = 0 \tag{3}$$

So:

$$|\phi\rangle = 0 \tag{4}$$

And the vector representing the sample space must be normalized, because:

$$\langle\Omega|\Omega\rangle = p(\Omega) = 1 \Rightarrow$$

$$\|\Omega\| = 1 \tag{5}$$

Furthermore, we know that the probability of an event is equal to the sum of the probabilities of the elementary events that constitute it[3], for example if:

$$A = \{a, b, ..., d\} \tag{6}$$

So:

$$p(A) = p(\{a\}) + p(\{b\}) + ... + p(\{d\}) \tag{7}$$

So, are (I), (II) and (III) consistent with this rule? Actually they are. To see that, let us suppose that sample space is:

$$\Omega = \{u_1, u_2, ..., u_N\} \tag{8}$$

And let us take some event $A \neq \phi$ to be:

$$A = \{u_k, ..., u_l\} \tag{9}$$

Let us take the orthonormal basis $\{|u_i\rangle\}_{i=1}^{N}$ to represent the elementary events such that:
$\{u_i\}$ is represented by

$$c_i|u_i\rangle : p(\{u_i\}) = |c_i|^2 = c_i^* c_i \tag{10}$$

Since $\{|u_i\rangle\}_{i=1}^{N}$ is a orthonormal basis, then it satisfies:

$$\sum_{i=1}^{N}|u_i\rangle\langle u_i| = I \tag{11}$$

$$\langle u_i|u_j\rangle = \delta_{ij} \tag{12}$$

Where $I$ is the identity operator.
According to (III), $A$ must be represented by:

$$|A\rangle = c_k|u_k\rangle + ... + c_l|u_l\rangle \tag{13}$$

And by adopting the notation:

$$|\alpha_k u_k + ... + \alpha_l u_l\rangle \equiv \alpha_k|u_k\rangle + ... + \alpha_l|u_l\rangle \tag{14}$$

$$\langle\alpha_k u_k + ... + \alpha_l u_l| \equiv \alpha_k^*\langle u_k| + ... + \alpha_l^*\langle u_l| \tag{15}$$

We have:

$$\langle A|A\rangle = \langle c_k u_k + ... + c_l u_l | c_k u_k + ... + c_l u_l\rangle$$

$$= |c_k|^2 + ... + |c_l|^2 = p_k + ... + p_l \tag{16}$$

Obviously, we see that $|\phi\rangle$ is represented using this basis as:

$$|\phi\rangle = |0\rangle = \sum_{i=1}^{N} 0|u_i\rangle \tag{17}$$

As a result of (II) and (III) we see that $\Omega$ is represented by:

$$|\Omega\rangle = \sum_{i=1}^{N} c_i|u_i\rangle \tag{18}$$

And we see that:

$$\langle\Omega|\Omega\rangle = \sum_{i=1}^{N}|c_i|^2 = \sum_{i=1}^{N} p_i = 1 = p(\Omega) \tag{19}$$

As an example that helps clarifying the former ideas, let us take the experiment to be throwing a fair die and the result to be the number appearing on top of it after it stabilizes on a horizontal surface.
The sample space of this experiment is:

$$\Omega = \{1,2,...,6\} \qquad (20)$$

Let us take $\{|i\rangle\}_{i=1}^{6}$ to be a orthonormal basis in Hilbert space, so:

$$\sum_{i=1}^{6}|i\rangle\langle i| = I \qquad (21)$$

$$\langle i|j\rangle = \delta_{ij} \qquad (22)$$

Let us choose to represent the elementary event $\{i\}$ by the vector:

$$\tfrac{1}{\sqrt{6}}|i\rangle \qquad (23)$$

That means $\Omega$ is represented by:

$$|\Omega\rangle = \sum_{i=1}^{6} \tfrac{1}{\sqrt{6}}|i\rangle \qquad (24)$$

and the event $A=\{2,5\}$ for example is represented by the vector:

$$|A\rangle = \tfrac{1}{\sqrt{6}}|2\rangle + \tfrac{1}{\sqrt{6}}|5\rangle \qquad (25)$$

but the event $\phi$ is represented by the zero vector.

**2.1 The algebra of events**

2.1.1 The intersection of two events

The intersection of two events is an event constituted of the common elements of the two events[3][ 4]. So, it must be represented by the vector sum of the vectors representing the common elementary events of the two vectors.

For example, in the die experiment mentioned above, if we took the two events: $A=\{1,2,3\}, B=\{2,3,4\}$ then $A\cap B=\{2,3\}$.

So:

$$|A\rangle = \tfrac{1}{\sqrt{6}}|1\rangle + \tfrac{1}{\sqrt{6}}|2\rangle + \tfrac{1}{\sqrt{6}}|3\rangle \qquad (26)$$

$$|B\rangle = \tfrac{1}{\sqrt{6}}|2\rangle + \tfrac{1}{\sqrt{6}}|3\rangle + \frac{1}{\sqrt{6}}|4\rangle \qquad (27)$$

And:

$$|A\cap B\rangle = \tfrac{1}{\sqrt{6}}|2\rangle + \tfrac{1}{\sqrt{6}}|3\rangle \qquad (28)$$

Now, we will prove a lemma. Supposing the sample space of some experiment is:

$$|\Omega\rangle = \sum_{i=1}^{N} c_i |u_i\rangle \qquad (29)$$

We see that for any event $A$, we have:

$A\cap\phi = \phi$ and $\langle A|\phi\rangle = \langle\phi|A\rangle = 0$ because $|\phi\rangle = 0$.

Now let us take two arbitrary events $A$ and $B$ $(A\neq\phi, B\neq\phi)$

And let us suppose they are represented by:

$$|A\rangle = c_k|u_k\rangle + ... + c_l|u_l\rangle \qquad (30)$$

$$|B\rangle = c_m|u_m\rangle + ... + c_n|u_n\rangle \qquad (31)$$

Then we have:

$$\langle A|B\rangle = \langle c_k u_k + ... + c_l u_l | c_m u_m + ... + c_n u_n\rangle$$

$$= c_k^* c_m \delta_{km} + ... + c_k^* c_n \delta_{kn} + ... + c_l^* c_m \delta_{lm} + ... + c_l^* c_n \delta_{ln} \qquad (32)$$

We notice that if $A\cap B = \phi$ then all of deltas are zeros so $\langle A|B\rangle = 0$.

So, we get this result:

$$A\cap B = \phi \Rightarrow \langle A|B\rangle = 0 \text{ so } |A\rangle \perp |B\rangle \qquad (33)$$

We see that for any two events $|A\rangle$ and $|B\rangle$, we can write the event $|A\rangle$ as:

$$|A\rangle = |A\cap B\rangle + |A_1\rangle : A_1 \cap (A\cap B) = \phi, A_1 \cup (A\cap B) = A \qquad (34)$$

But since $A_1 \cap (A\cap B) = \phi$ then as we see according to (33) that:

$$\langle A_1 | A\cap B\rangle = 0 \qquad (35)$$

So:

$$\langle A\cap B|A\rangle = \langle A\cap B|(|A_1\rangle + |A\cap B\rangle)$$
$$= \langle A\cap B|A_1\rangle + \langle A\cap B|A\cap B\rangle$$
$$= \langle A\cap B|A\cap B\rangle \qquad (36)$$

So:

$$\langle A\cap B|A\rangle = \langle A\cap B|A\cap B\rangle \qquad (37)$$

And:

$$\langle A\cap B|B\rangle = \langle A\cap B|A\cap B\rangle \qquad (38)$$

As a result:

$$\langle A\cap \Omega|\Omega\rangle = \langle A\cap \Omega|A\cap \Omega\rangle \qquad (39)$$

But since $A\cap\Omega = A$, so:

$$\langle A|\Omega\rangle = \langle A|A\rangle \qquad (40)$$

Or:

$$\langle A|\Omega\rangle = p(A) \qquad (41)$$

So:
$$\langle A|\Omega\rangle \geq 0 \qquad (42)$$

$$\langle A|\Omega\rangle = \langle\Omega|A\rangle = \langle A|A\rangle = p(A) \qquad (43)$$

We can write the former results, since the probability of some event is equal to the square of its norm, and using (37), by the following manner:

$$\langle A\cap B|A\cap B\rangle = \langle A\cap B|A\rangle \Rightarrow$$

$$p(A\cap B) = \langle A\cap B|A\rangle \qquad (44)$$

And:
$$\langle A\cap B|A\cap B\rangle = \langle A\cap B|A\rangle \Rightarrow$$
$$\langle A\cap B|A\cap B\rangle^* = \langle A|A\cap B\rangle \Rightarrow$$
$$\langle A\cap B|A\cap B\rangle = \langle A|A\cap B\rangle \Rightarrow$$

$$\langle A\cap B|A\rangle = \langle A|A\cap B\rangle \qquad (45)$$

Finally, we can see also that, if we have two events:

$$|A\rangle = c_k|u_k\rangle + \ldots + c_l|u_l\rangle \qquad (46)$$

$$|B\rangle = c_m|u_m\rangle + \ldots + c_n|u_n\rangle \qquad (47)$$

Then we can write the intersection of them as:
$$|A\cap B\rangle = |B\cap A\rangle = \langle u_k|B\rangle|u_k\rangle + \ldots + \langle u_l|B\rangle|u_l\rangle$$
$$= \langle u_m|A\rangle|u_m\rangle + \ldots + \langle u_n|A\rangle|u_n\rangle \qquad (48)$$

### 2.1.2 The difference of two events

Let us take two events $|A\rangle$ and $|B\rangle$. We saw from (34) that we can write the event $|A\rangle$ as:

$$|A\rangle = |A_1\rangle + |A\cap B\rangle$$

Where $|A_1\rangle$ is an event constituted of elements not present in $|B\rangle$ but belong to $|A\rangle$ so:

$$|A\setminus B\rangle = |A_1\rangle$$

That means:

$$|A\rangle = |A\setminus B\rangle + |A\cap B\rangle$$

So:
$$|A\setminus B\rangle = |A\rangle - |A\cap B\rangle \qquad (49)$$

We can verify immediately that:
$$\langle A\setminus B|A\setminus B\rangle = \langle(A)-(A\cap B)|(A)-(A\cap B)\rangle$$
$$= \langle(A)-(A\cap B)|A\rangle - \langle(A)-(A\cap B)|A\cap B\rangle$$
$$= \langle A|A\rangle - \langle A\cap B|A\rangle - \langle A|A\cap B\rangle + \langle A\cap B|A\cap B\rangle$$
$$= p(A) - p(A\cap B) - p(A\cap B) + p(A\cap B)$$
$$= p(A) - p(A\cap B) = p(A\setminus B)$$

### 2.1.3. The union of two events

We know that the union of two events is an event constituted of the elements belonging exclusively to the first one, the elements belonging exclusively to the second one and the common elements between the two[3][4].

So, it must be represented by:

$$|A\cup B\rangle = |A\setminus B\rangle + |A\cap B\rangle + |B\setminus A\rangle$$

Which we can write as:
$$|A\cup B\rangle = |A\setminus B\rangle + |A\cap B\rangle + |B\setminus A\rangle + |A\cap B\rangle - |A\cap B\rangle \Rightarrow$$

$$|A\cup B\rangle = |A\rangle + |B\rangle - |A\cap B\rangle \qquad (50)$$

Noting that:
$$\langle B|A\setminus B\rangle = 0$$

Because $B\cap(A\setminus B) = \phi$ we can directly verify that:
$$\langle A\cup B|A\cup B\rangle = \langle(A)+(B)-(A\cap B)|(A)+(B)-(A\cap B)\rangle$$
$$= p(A) + p(B) - p(A\cap B) = p(A\cup B)$$

In a side note, we can prove that:
$$\langle A|B\rangle = \langle A|(|B\setminus A\rangle + |A\cap B\rangle) = \langle A|B\setminus A\rangle + \langle A|A\cap B\rangle$$
$$= 0 + \langle A|A\cap B\rangle = \langle A|A\cap B\rangle$$

So:
$$\langle A|B\rangle = \langle B|A\rangle = \langle A|A\cap B\rangle = \langle B|A\cap B\rangle = p(A\cap B) \qquad (51)$$

### 2.1.4. The complementary event

We know that the complementary event $A'$ of an event $A$ is given by[3][4]:

$$A' = \Omega\setminus A$$

So:
$$|A'\rangle = |\Omega\setminus A\rangle = |\Omega\rangle - |\Omega\cap A\rangle = |\Omega\rangle - |A\rangle$$

So we have:

$$|A'\rangle = |\Omega\rangle - |A\rangle \qquad (52)$$

We can directly verify that:

$$\langle A'|A'\rangle = \langle \Omega - A|\Omega - A\rangle = 1 - p(A) = p(A')$$

# 3. Observables

Let us suppose we have a system, and we want to do an experiment with it, which has the sample space:

$$\Omega = \{u_1, ..., u_N\} \qquad (53)$$

Or equivalently:

$$|\Omega\rangle = \sum_{i=1}^{N} c_i |u_i\rangle \qquad (54)$$

Where as we saw, since $\{|u_i\rangle\}_{i=1}^{N}$ is a orthonormal basis:

$$\sum_{i=1}^{N} |u_i\rangle\langle u_i| = I \text{ and } \langle u_i|u_j\rangle = \delta_{ij}$$

Now, if we take any $N$ real numbers:

$$\lambda_1, \lambda_2, ..., \lambda_N$$

Then we can consider them to be the eignvalues of a Hermitian operator $\hat{A}$ which is represented in this basis by the matrix:

$$diag(\lambda_1, ..., \lambda_N)$$

Clearly, the vectors of the ordered basis $\{|u_i\rangle\}_{i=1}^{N}$ are the eigenvectors of $\hat{A}$ which satisfy:

$$\hat{A}|u_i\rangle = \lambda_i |u_i\rangle \qquad (55)$$

From the fact that this is true for any lambdas, in other words the values of $\lambda_i$ are arbitrary, then the vectors of the basis $\{|u_i\rangle\}_{i=1}^{N}$ are the eigenvectors of an infinite number of Hermitian operators in Hilbert space.

Not even just that, but since this is true for any lambdas, then whenever we assign real numbers to elementary events, we can consider them to be the eignvalues of some Hermitian operator in Hilbert space corresponding to the eigenvectors $\{|u_i\rangle\}_{i=1}^{N}$.

And since the observable is by definition a function from the elementary events to real numbers[1], then we can represent any observable we define on the system, by a Hermitian operator in Hilbert space.

But we have to be careful here: all the observables we have talked about have the same set of eigenvectors, and we will call them compatible observables, and if we take any two of them, we find that their commutator is zero, because they have the same eigenvectors.

If we take one of these observables, let it be $\hat{A}$, which is represented by:

$$diag(\lambda_1, ..., \lambda_N)$$

Then we can think of the experiment as giving us one eignvalue of the observable. And since this is true for every one of the compatible observables with $\hat{A}$ as we saw, then it is clear that compatible observables can be measured simultaneously together with a single experiment, which is the experiment we talked about.

Now, let us suppose that:

$$\underbrace{\lambda_i = \lambda_j = ... = \lambda_k}_{g \text{ values}} = \lambda \qquad (56)$$

Which means that $\lambda$ is degenerate with a degeneracy $g$. That means we will get $\lambda$ in the experiment if we get $u_i$, $u_j$,..., or $u_k$. In other words, if the event: $B = \{u_i, u_j, ..., u_k\}$ happened.

So:

$$p(\lambda) = p(B) = \langle B|B\rangle \Rightarrow$$
$$p(\lambda) = \langle c_i u_i + c_j u_j + ... + c_k u_k | c_i u_i + c_j u_j + ... + c_k u_k \rangle$$
$$= |c_i|^2 + |c_j|^2 + ... + |c_k|^2 \Rightarrow$$

$$p(\lambda) = |\langle u_i|\Omega\rangle|^2 + |\langle u_j|\Omega\rangle|^2 + ... + |\langle u_k|\Omega\rangle|^2 \qquad (57)$$

So we see, that if $\lambda$ was degenerate, then it is probability is equal to the probability of the projection of $|\Omega\rangle$ on the subspace spanned by the eigenvectors corresponding to $\lambda$.

But what about other observables we can define on the system corresponding to other experiments? Those experiments may have in general totally different probability distributions, which means the probabilities of their elementary events are different of those in the experiment we have talked about in the beginning of this section. Even more, the number of the elementary events may be different.

Let us suppose we have a system. And let us assume that we intend to do some experiment on the system which has the sample space:

$$\Omega = \{u_1, ..., u_k\} \qquad (58)$$

Or equivalently:

$$|\Omega\rangle = \sum_{i=1}^{N} c_i |u_i\rangle \qquad (59)$$

Now we will divide all the experiments we can do on the system into classes of experiments. Each class is composed of experiments that have the same number of outcomes (the same number of elementary events). So the experiment that we talked about is one member of the class $C_N$ where $C_N$ is the class of experiments that have $N$ elementary events by definition.

We will name our experiment $E_1$. We saw that for the experiment $E_1$, we can define an infinite number of compatible observables (Hermitian operators) which all have the same eigenvectors $\{|u_i\rangle\}_{i=1}^{N}$. Let us now take another experiment from the same class $C_N$ which we will call $E_2$. What we mean by another experiment on the system is that we cannot do $E_1$ and $E_2$ simultaneously.

Let us suppose the sample space of $E_2$ is:

$$\Gamma = \{t_1, t_2, ..., t_N\} \tag{60}$$

Now in general, the probability distribution of $E_2$ may be radically different from that of $E_1$. So, how are we going to represent the events of $E_2$ by vectors? Well, since we can represent $\Gamma$ by a vector in any N-dimensional Hilbert space, then we can represent it in the same Hilbert space that we used to represent $\Omega$. We can use another basis in this space and use another vector (different from $|\Omega\rangle$) to represent $\Gamma$, or we can use the same basis and a different vector from $|\Omega\rangle$ to represent $\Gamma$, or we can use a different basis (different from $\{|u_i\rangle\}_{i=1}^{N}$) and the same vector $|\Omega\rangle$ to represent $\Gamma$, where the components of $|\Omega\rangle$ on the new basis are which determine the probabilities of the events of $E_2$. All these approaches are valid, but we will choose the last one (we could also have worked in a different Hilbert space all together). Of course we can represent $E_2$ with the same basis and the same vector for the sample space, if it has the same probability distribution of $E_1$. But to distinguish $E_2$ as an experiment that cannot be done simultaneously with $E_1$, we will represent its elementary events by a different basis.

Now for $E_2$, we have:

$$|\Omega\rangle = \sum_{j=1}^{N} b_j |t_j\rangle \tag{61}$$

As in the case of $E_1$, here, we can define an infinite number of observables (Hermitian operators), all compatible, and which have the eigenvectors $\{|t_j\rangle\}_{j=1}^{N}$, and those observables are associated with $E_2$. Since all of them have the same set of eigenvectors, then the commutator of any two of them is zero.

But if we take one of them, let it be $\hat{B}$, and take an observable $\hat{A}$ associated with $E_1$, then since $\hat{A}$ and $\hat{B}$ do not have the same eigenvectors (because $\{|u_i\rangle\}_{i=1}^{N}$ is not $\{|t_j\rangle\}_{j=1}^{N}$), then $[\hat{A}, \hat{B}] \neq 0$.

And since $E_1$ and $E_2$ cannot be done simultaneously, then we cannot measure $\hat{A}$ and $\hat{B}$ simultaneously, because each observable is defined in terms of the experiment it is associated with. So we call them incompatible.

Now, before we continue, let us take some examples of some compatible and incompatible observables.

1- Compatible observables:

In the experiment of throwing the die, we can define the first observable to be the number appearing on the top side of the die, and the second observable to be the square of the number appearing on the top side of the die.

Let us call the first $\hat{A}$ and the second $\hat{B}$. We have:

$$|\Omega\rangle = \sum_{i=1}^{6} \tfrac{1}{\sqrt{6}} |i\rangle \tag{62}$$

Where:

$$\langle i|j\rangle = \delta_{ij} \text{ and } \sum_{i}^{6} |i\rangle\langle i| = I \tag{63}$$

We have:

$$\hat{A}|i\rangle = i|i\rangle \tag{64}$$

And:

$$\hat{B}|i\rangle = i^2 |i\rangle \tag{65}$$

So $\hat{A}$ is represented by:

$$diag(1,2,3.4,5,6) \tag{66}$$

While $\hat{B}$ is represented by:

$$diag(1,4,9,16,25,36) \tag{67}$$

We see that:

$$[\hat{A}, \hat{B}] = \hat{A}\hat{B} - \hat{B}\hat{A} = 0 \tag{68}$$

2- Incompatible observables:

Let us take a coin. We will imagine two ideal experiments that we can do with it. In the first one, let us call it $E_1$, we toss the coin and it stabilizes on a horizontal surface and the top side of it is either Heads or Tails. We can define the observable $\hat{A}$ to take the value 1 for Heads, and the value

-1 for Tails. The second experiment, $E_2$, is throwing the coin in a particular way, that makes it stabilize on its edge on some horizontal surface. We suppose that the edge of the coin is half painted. We can define an observable $\hat{B}$ to take the value 1 if we looked at the coin from above and saw the edge either all painted or all not painted, and -1 if we saw it partially painted. We see that we cannot do both $E_1$ and $E_2$ simultaneously, so: $[\hat{A},\hat{B}] \neq 0$.

*(Notice that for some class of experiments $C_N$, once we used a vector to represent the sample space of some experiment, then we have to ask ourselves, can we use it to represent all the sample spaces of all the experiments of this class that can be done on the system?

Clearly, when we want to represent the sample space of some experiment by a vector in Hilbert space, we can choose any Hilbert space that has the right dimensionality, and any orthonormal basis in it to represent the elementary events, and a vector to represent the sample space that satisfies that the squared norms of its components are the probabilities of elementary events. But after this if we want this vector to represent all the experiments of this class, we have to choose the bases representing those experiments carefully. Or we can choose the bases that represent all experiments in Hilbert space, then look for the vector that can be used as a sample space vector for all of them.

And as we will see in the future, not every vector we use to represent the sample space of some experiment of some class, satisfies this for all experiments of the given class. So, we will call any vector that actually satisfies this condition, meaning it represents the sample space of all possible experiments of a given class that can be done on the system, we will call it the state vector of the system because it gives us the information about any experiment we can do on the system for a given class of experiments. And from now on, throughout this paper, when we use the term "state vector", we mean it in this particular sense. We will talk more about this later).

Now, what if we take two experiments from different classes, say $C_N$ and $C_M$ where $M \neq N$?

Let us suppose that the experiment $E_1$ is from the class $C_N$ and $E_2$ is from the class $C_M$. Here we will represent each experiment in its own Hilbert space: $E_1$ in $H_1$ and $E_2$ in $H_2$.

Let us suppose we represent the sample space of $E_1$ in $H_1$ by:

$$|\Omega_1\rangle = \sum_{i=1}^{N} c_i |u_i\rangle \in H_1 \quad (69)$$

And that we represent the sample space of $E_2$ in $H_2$ by:

$$|\Omega_2\rangle = \sum_{j=1}^{M} b_j |t_j\rangle \in H_2 \quad (70)$$

Now, let us suppose that we do both experiments on the system. Even more, we will suppose that doing either experiment does not affect the probability distribution of the other, whatever the order of doing the two experiments was.

When we do $E_1$, we have the sample space:

$$\Omega_1 = \{u_1,...,u_N\} \quad (71)$$

Let us suppose that after doing $E_1$ we do $E_2$ and that for $E_2$ the sample space is:

$$\Omega_2 = \{t_1,...,t_M\} \quad (72)$$

Let us take the composite experiment $E$ that is doing $E_1$ then $E_2$ on the system. The sample space of this experiment is:

$$\Omega = \{\underbrace{(u_1,t_1),(u_1,t_2),...,(u_N,t_M)}_{N \cdot M}\} \quad (73)$$

Where we know that, if the probability of $\{u_i\}$ in $E_1$ is $p(u_i)$ and the probability of $\{t_j\}$ in $E_2$ is $p(t_j)$ then the probability of $\{(u_i,t_j)\}$ is[4]:

$$p = p(u_i).p(t_j) \quad (74)$$

That is interesting, because if we take the vector space:

$$H = H_1 \otimes H_2 \quad (75)$$

and take the vector $|\Omega\rangle$ in it which is:

$$|\Omega\rangle = |\Omega_1\rangle \otimes |\Omega_2\rangle = \left(\sum_{i=1}^{N} c_i |u_i\rangle\right) \otimes \left(\sum_{j=1}^{M} b_j |t_j\rangle\right)$$
$$= \sum_{i,j}^{N \cdot M} c_i b_j |u_i\rangle \otimes |t_j\rangle$$
$$\equiv \sum_{i,j}^{N \cdot M} c_i b_j |u_i t_j\rangle \quad (76)$$

Where by definition:

$$|u_i t_j\rangle \equiv |u_i\rangle \otimes |t_j\rangle \quad (77)$$

First of all, we see that the dimensionality of $H$ is $(N.M)$. If we considered $|\Omega\rangle$ to be the vector representing the sample space of some experiment that has $(N.M)$ elementary events, then the probability of the elementary event

$|u_i t_j\rangle$ is:

$$p = |c_i b_j|^2 = |c_i|^2 |b_j|^2 = p(u_i) p(t_j) \quad (78)$$

Well, it is the same probability of the event $\{(u_i, t_j)\}$ in the experiment $E$ which has a $(N.M)$ elementary events. From the above we see that we can represent the sample space of the composite experiment $E$ by a vector $|\Omega\rangle$ in the Hilbert space $H_1 \otimes H_2$ where:

$$|\Omega\rangle = |\Omega_1\rangle \otimes |\Omega_2\rangle \quad (79)$$

Where the elementary event $\{(u_i, t_j)\}$ is represented by $c_i b_j |u_i t_j\rangle$.

Now, what if doing one experiment affects the probability distribution of the other? Here, we can still write the sample space of the composite experiment as:

$$\begin{aligned}\Omega &= \Omega_1 \times \Omega_2 \\ &= \{(u_1,t_1),...,(u_1,t_M),...,(u_N,t_1),...,(u_N,t_M)\}\end{aligned} \quad (80)$$

Because the outcomes of the two experiments remain the same, but what is changing, is probability distributions. So, we cannot write: $p(\{u_i, t_j\}) = p\{u_i\} p\{t_j\}$ because finding $u_i$ is not independent from finding $t_j$ in general.

Still, the outcomes of the composite experiment are $N \cdot M$ elementary events, so we have to represent $\Omega$ in a Hilbert space with $N \cdot M$ dimensionality. And any Hilbert space with this dimensionality will do. So, we can represent $\Omega$ in:

$$H = H_1 \otimes H_2$$

And since we can choose any orthonormal basis in it, and since $\{|u_i t_j\rangle\}_{i=1}^{N \cdot M}$ are orthonormal basis, so we can represent $\Omega$ by:

$$|\Omega\rangle = \sum_{i,j} f_{i,j} |u_i t_j\rangle \in H_1 \otimes H_2 = H \quad (81)$$

Where $\langle \Omega | \Omega \rangle = 1$ and $f_{i,j} |u_i t_j\rangle$ representing the elementary event $\{(u_i, t_j)\}$, thus:

$$p(\{(u_i, t_j)\}) = |f_{i,j}|^2 \quad (82)$$

We can generalize this to any number of experiments.

P.S. when we define the sample space of some experiment, it is not necessary that we really do the experiment, but it just describes a potential experiment. We will use this later. Now, let us ask ourselves a question: is the state vector unique? Can we use for a given class of experiments, more than one vector as a state vector?

If it is not unique, then we must find the same probabilities for all experiments of this class that we can do on the system, whether we used $|\Omega\rangle$ or –if exist- the other vector/vectors that can be used as state vectors.

Let us suppose that for an experiment $E_1$, the state vector of the system is written as:

$$|\Omega\rangle = \sum_{l=1}^{N} c_l |u_l\rangle \quad (83)$$

Let us take the vector $|\Omega'\rangle$ which is:

$$|\Omega'\rangle = \sum_{l=1}^{N} c'_l |u_l\rangle : c'_l = c_l z_l \quad (84)$$

Where $z_l$ are complex numbers which we will write in the form:

$$z_l = A_l e^{i\theta_l} : A_l \in R^+, \theta_l \in R \quad (85)$$

Where: $i = \sqrt{-1}$.

For $|\Omega'\rangle$ to be a state vector for the system, then all the probabilities of the elementary events (so all the probabilities of all events since the probability of an event is the sum of the probabilities of its elementary events) of any experiment from this class must be the same as given by $|\Omega\rangle$. So, the probabilities of the elementary events of $E_1$ do not change.

So the following equation must hold:

$$|c'_l|^2 = |c_l|^2 \quad (86)$$

So:

$$|A_l e^{i\theta_l} c_l|^2 = |c_l|^2 \Rightarrow$$

$$|A_l|^2 |c_l|^2 = |c_l|^2 \quad (87)$$

And because the former condition is true even if we choose the experiment to satisfy: $c_l \neq 0$ for all values of $c_l$, because our choice of $E_1$ is arbitrary, we must have:

$$|A_l|^2 = 1$$

So we have the condition:
$$A_l = 1 \tag{88}$$

Which means that:
$$z_l = e^{i\theta_l} \tag{89}$$

And that:
$$c'_l = c_l z_l = c_l e^{i\theta_l} \tag{90}$$

But that is not enough, because the condition that probabilities must not change must be true for any other experiment from the same class, that we can do on the system and not just $E_1$, because we are talking about state vectors here. Let us take another experiment $E_2$ of the same class. We know that it must be represented by another basis, let us say $\{|t_j\rangle\}_{j=1}^N$. We must have:

$$|\Omega\rangle = \sum_{j=1}^N b_j |t_j\rangle \tag{91}$$

We have:
$$|\Omega\rangle = \sum_{j=1}^N b_j |t_j\rangle = \sum_{l=1}^N c_l |u_l\rangle \tag{92}$$

Where:
$$b_j = \langle t_j|\Omega\rangle = \sum_{l=1}^N \langle t_j|u_l\rangle\langle u_l|\Omega\rangle = \sum_{l=1}^N \langle t_j|u_l\rangle c_l \tag{93}$$

For $|\Omega'\rangle$ we must have:
$$|\Omega'\rangle = \sum_{j=1}^N b'_j |t_j\rangle \tag{94}$$

So:
$$b'_j = \langle t_j|\Omega'\rangle = \sum_{l=1}^N \langle t_j|u_l\rangle\langle u_l|\Omega'\rangle$$
$$= \sum_{l=1}^N \langle t_j|u_l\rangle c'_l = \sum_{l=1}^N \langle t_j|u_l\rangle c_l e^{i\theta_l} \tag{95}$$

We saw that the probabilities of the events of $E_1$ do not change. But to reach our goal, which is that we want $|\Omega'\rangle$ to be a state vector too, then the probabilities of the events of $E_2$ must not change. So, we must have:

$$|b'_j|^2 = |b_j|^2$$

$$b'_j b'^*_j = b_j b^*_j$$

$$\left(\sum_{l=1}^N \langle t_j|u_l\rangle c_l e^{i\theta_l}\right)\left(\sum_{k=1}^N \langle u_k|t_j\rangle c_k^* e^{-i\theta_k}\right) = \left(\sum_{l=1}^N \langle t_j|u_l\rangle c_l\right)\left(\sum_{k=1}^N \langle u_k|t_j\rangle c_k^*\right)$$

So we must have:

$$\sum_{l,k} \langle t_j|u_l\rangle\langle u_k|t_j\rangle c_l c_k^* e^{i(\theta_l-\theta_k)} = \sum_{l,k} \langle t_j|u_l\rangle\langle u_k|t_j\rangle c_l c_k^* \tag{96}$$

The former equation must be true for any $c_k$ and $c_l$ because we are speaking of arbitrary experiments with arbitrary probability distributions. It is true when we fix the basis whatever $c_l$ were. So their coefficients must be the same (we fixed the two bases and can define an infinite number of experiments on them):

$$\langle t_j|u_l\rangle\langle u_k|t_j\rangle e^{i(\theta_l-\theta_k)} = \langle t_j|u_l\rangle\langle u_k|t_j\rangle \Rightarrow$$
$$\langle t_j|u_l\rangle\langle u_k|t_j\rangle[e^{i(\theta_l-\theta_k)} - 1] = 0 \tag{97}$$

It must be true for all experiments so for all bases even when: $\langle t_j|u_l\rangle \neq 0$ for any $l$ and $j$. So:

$$e^{i(\theta_l-\theta_k)} = 1$$

So:
$$e^{i\theta_l} = e^{i\theta_k} \Rightarrow \theta_l = \theta_k \tag{98}$$

And that is for any $l$ and $k$. So we have:
$$z_l = z_k \tag{99}$$

So we see that:
$$z_1 = z_2 = ... = z_N \equiv e^{i\theta} \tag{100}$$

So:
$$|\Omega'\rangle = \sum_{l=1}^N c_l e^{i\theta}|u_l\rangle = e^{i\theta}\sum_{l=1}^N c_l|u_l\rangle \Rightarrow$$
$$|\Omega'\rangle = e^{i\theta}|\Omega\rangle \tag{101}$$

So, for $|\Omega'\rangle$ to be a state vector too, it must be of the former form. From the above we see that we can multiply $|\Omega\rangle$ by any pure phase and still get another state vector.

## 4. The collapse of the state vector

Let us suppose that we have a system. We want to do on it an experiment $E_1$ of the class $C_N$. That means the state vector of it is:

$$|\Omega\rangle = \sum_{i=1}^N c_i |u_i\rangle \tag{102}$$

Let us suppose that the result of the experiment was $u_k$, which means that the elementary event $\{u_k\}$ has happened. Let us suppose that we want to do another experiment now on the system from the same class, after we did the first one.

Well, one such experiment could be just reading the result of the former experiment. Since the result was $u_k$ then definitely we will find the result $u_k$. So we can represent the sample space of this experiment by:

$$\sum_{i=1}^{N}\delta_{ik}|u_i\rangle = |u_k\rangle \qquad (103)$$

But, according to the note (*), the state vector after the measurement i.e. after doing the second experiment (after reading the result), must be able to represent this experiment. So it must give the same probabilities for the elementary events of this experiment. So we must have:

$$p(\{u_l\}) = \delta_{kl} \qquad (104)$$

Which means the state vector after the measurement(i.e. after doing the experiment) must be of the form:

$$|\Omega'\rangle = e^{i\theta}|u_k\rangle = e^{i\theta}\sum_{i=1}^{N}\delta_{ik}|u_i\rangle \qquad (105)$$

we can use this vector to represent any other sample space of any experiment of the same class $C_N$, which we do after the measurement, because it is the state vector after the measurement.

So, if the state vector before we do some experiment was given by (102), and after that, we did the experiment and got the result $u_k$, then the state vector of class $C_N$ after the measurement becomes:

$$|\Omega'\rangle = e^{i\theta}|u_k\rangle \qquad (106)$$

Of course we can choose any value for $\theta$ including $0$ so we can write the state vector after the measurement as:

$$|\Omega'\rangle = |u_k\rangle \qquad (107)$$

We can call this a collapse in the state vector. But we also see that there is nothing mysterious here, for we just have a change in probability distribution after the measurement.

We can see that another way to express the above is, that if $\hat{A}$ is an observable that the experiment measures (as we have mentioned, that means a Hermitian operator that has $\{|u_i\rangle\}_{i=1}^{N}$ as its eigenvectors), then the system after the measurement will be in an eigen state of $\hat{A}$ corresponding to the eigenvalue of it that we will measure.

## 5. Entangled states

Now, how do we represent composite systems?
Let us at first take two non-interacting systems. Let us do an experiment $E_1$ on the first one of the class $C_{1N}$ ($C_{1N}$ is the class of experiments we can do on the first system with $N$ outcomes). Its sample space will be of the form:

$$\Omega_1 = \{u_1,...,u_N\} \qquad (108)$$

and we will suppose the state vector of it is:

$$|\Omega_1\rangle = \sum_{i=1}^{N}c_i|u_i\rangle \in H_1 \qquad (109)$$

And let us suppose that the sample space of the second experiment $E_2$ on the second system of the class $C_{2M}$ is:

$$\Omega_2 = \{t_1,...,t_M\} \qquad (110)$$

and we will suppose its state vector is

$$|\Omega_2\rangle = \sum_{j=1}^{M}b_j|t_j\rangle \in H_2 \qquad (111)$$

We know that the sample space of the composite experiment $E$ of doing $E_1$ and $E_2$ together is:

$$\Omega = \Omega_1 \times \Omega_2$$
$$= \{(u_1,t_1),...,(u_1,t_M),...,(u_N,t_1),...,(u_N,t_M)\} \qquad (112)$$

We see that $E$ has $N.M$ outcomes.

Even more, since the two systems do not interact with each other, then the probability of the result $(u_i,t_j)$ is:

$$p\{(u_i,t_j)\} = p\{u_i\}.p\{t_j\} \qquad (113)$$

Where $p\{u_i\}$ is the probability of getting $u_i$ in the experiment $E_1$, while $p\{t_j\}$ is the probability of getting $t_j$ in the experiment $E_2$.

We know that we can represent $\Omega$ in any Hilbert space of dimensionality $N.M$ by a vector that gives the elementary events of $E$ the former probabilities.

If we take the vector:

$$|\psi\rangle = |\Omega_1\rangle \otimes |\Omega_2\rangle \in H = H_1 \otimes H_2 \qquad (114)$$

We see that $H$ has the dimensionality $N.M$. Furthermore:

$$|\psi\rangle = |\Omega_1\rangle \otimes |\Omega_2\rangle = \left(\sum_{i=1}^{N}c_i|u_i\rangle\right) \otimes \left(\sum_{j=1}^{M}b_j|t_j\rangle\right)$$
$$= \sum_{i,j}^{N.M}c_ib_j|u_i\rangle \otimes |t_j\rangle \equiv \sum_{i,j}^{N.M}c_ib_j|u_it_j\rangle \qquad (115)$$

We see that:

$$\langle\psi|\psi\rangle = \sum_{i,j}|c_ib_j|^2 = \sum_{i,j}|c_i|^2|b_j|^2 \Rightarrow$$

$$\langle\psi|\psi\rangle = \sum_{i,j}p\{u_i\}.p\{t_j\} = \sum_{i,j}p\{(u_i,t_j)\} = 1 \qquad (116)$$

So, we can use $|\psi\rangle$ as a representation of the sample space of some experiment with $N.M$ outcomes. Furthermore, the basis vector $|u_i t_j\rangle$ can represent an elementary event with a probability:

$$p = |c_i b_j|^2 = |c_i|^2 |b_j|^2 = p\{u_i\}.p\{t_j\} = p\{(u_i,t_j)\} \quad (117)$$

From all that we see that we can use:

$$|\psi\rangle = |\Omega\rangle = |\Omega_1\rangle \otimes |\Omega_2\rangle \in H = H_1 \otimes H_2 \quad (118)$$

to represent $\Omega$ with $c_i b_j |u_i t_j\rangle$ representing the elementary event $\{(u_i, t_j)\}$.

And since we are working in a new space altogether, we can use this vector to be the state vector of the composite system, taking into account that we have to be careful after it, to choose the bases representing other experiments of the class $C_{N.M}$ on the composite system in the right way.

We can call $|\psi\rangle$ a product state.

We can take as an example of the above two non-interacting fair coins. The sample space of the experiment of tossing the first coin is:

$$\Omega_1 = \{H,T\} \quad (119)$$

and let us suppose that its state vector before measurement was:

$$|\Omega_1\rangle = \frac{1}{\sqrt{2}}|H\rangle + \frac{1}{\sqrt{2}}|T\rangle \in H_1 \quad (120)$$

And for the second coin:

$$\Omega_2 = \{H,T\} \quad (121)$$

And we suppose too that its state vector:

$$|\Omega_2\rangle = \frac{1}{\sqrt{2}}|H\rangle + \frac{1}{\sqrt{2}}|T\rangle \in H_2 \quad (122)$$

And the sample space of the composite system (tossing the two coins together) is:

$$\Omega = \{(H,H),(H,T),(T,H),(T,T)\} \quad (123)$$

So the state vector for the composite system of the two coins is:

$$|\Omega\rangle = |\Omega_1\rangle \otimes |\Omega_2\rangle$$
$$= \frac{1}{2}|HH\rangle + \frac{1}{2}|HT\rangle + \frac{1}{2}|TH\rangle + \frac{1}{2}|TT\rangle \quad (124)$$

Now: what if the two systems were interacting with each other?

Here, we can still write the sample space as given by the equation (112), and that is if the outcomes of the two experiments remain the same, but what interaction is changing, is probability distributions. So, we cannot write: $p\{(u_i,t_j)\} = p\{u_i\}.p\{t_j\}$ because finding $u_i$ is not in general independent from finding $t_j$ and vice versa..

Still, the outcomes of the composite experiment are $N.M$ elementary events, so we have to represent $\Omega$ in a Hilbert space with $N.M$ dimensionality. And any Hilbert space with this dimensionality will do. So, we can represent $\Omega$ in:

$$H = H_1 \otimes H_2 \quad (125)$$

And since we can choose any orthonormal basis in it, and since $\{|u_i t_j\rangle\}$ are orthonormal basis, and we are working in a whole new space $H = H_1 \otimes H_2$ different from both $H_1$ and $H_2$, thus we can represent the state vector of the composite system by:

$$|\Omega\rangle = \sum_{i,j} f_{i,j} |u_i t_j\rangle \in H_1 \otimes H_2 = H \quad (126)$$

Where $\langle\Omega|\Omega\rangle = 1$ and $f_{i,j}|u_i t_j\rangle$ representing the elementary event $\{(u_i, t_j)\}$.

We see that the product state is a special case of the former formula when $f_{i,j} = c_i b_j$.

We call any state $|\Omega\rangle = \sum_{i,j} f_{i,j} |u_i t_j\rangle$ an entangled state if it is not a product state.

Now we can talk about the composite (system-observer) system.

## 6. The observer-system composite system

Let us start by an example, then generalize. Let us take the composite system of (coin-coin tosser). The sample space of the coin is:

$$\Omega_1 = \{H,T\} \quad (127)$$

And we will suppose its state vector is:

$$|\Omega_1\rangle = \frac{1}{\sqrt{2}}|H\rangle + \frac{1}{\sqrt{2}}|T\rangle \quad (128)$$

The observer (coin tosser) may get two results: observer seeing the coin Heads, or observer seeing the coin Tails. So, the sample space of the experiment which is the observer observing the coin is:

$$\Omega_2 = \{O_H, O_T\} \quad (129)$$

And let us suppose its state vector is:

$$|\Omega_2\rangle = c_H|O_H\rangle + c_T|O_T\rangle \quad (130)$$

Bearing in mind that:

$$p\{O_H\} = p\{H\} \quad (131)$$

And that:

$$p\{O_T\} = p\{T\} \quad (132)$$

So, we have:

$$|c_H|^2 = |c_T|^2 = \frac{1}{2} \quad (133)$$

The state vector of the composite system (which can be thought of as the sample space of the experiment of observing the composite system of coin – coin tosser) as we know, can be written as:

$$|\Omega\rangle = c_1|HO_H\rangle + c_2|HO_T\rangle + c_3|TO_H\rangle + c_4|TO_T\rangle \quad (134)$$

But we know that we can only get either $|HO_H\rangle$ or $|TO_T\rangle$, so:

$$c_2 = c_3 = 0 \quad (135)$$

Even more, the probability of the observer sees it Heads is the same as the probability of getting Heads. The same goes for tails. So:

$$|c_1|^2 = |c_4|^2 = \frac{1}{2} \quad (136)$$

So:

$$|\Omega\rangle = c_1|HO_H\rangle + c_4|TO_T\rangle \neq |\Omega_1\rangle \otimes |\Omega_2\rangle \quad (137)$$

So, $|\Omega\rangle$ is an entangled state. In the same way, if we have a system with a state vector:

$$|\Omega_1\rangle = \sum_{i=1}^{N} c_i|u_i\rangle \in H_1 \quad (138)$$

Then each $c_i|u_i\rangle$ corresponds to an elementary event of the observer which has the same probability, so it can be represented by $b_i|O_i\rangle$ where:

$$|b_i|^2 = |c_i|^2 \quad (139)$$

and the state vector of the observer is:

$$|\Omega_2\rangle = \sum_{i=1}^{N} b_i|O_i\rangle \in H_2 \quad (140)$$

And the state vector of the composite system (which is the sample space of the experiment that is observing the composite system) is of the form:

$$|\Omega\rangle = \sum_{i,j} d_{i,j}|u_iO_j\rangle \in H_1 \otimes H_2 \quad (141)$$

But since the probability of the event $d_{i,j}|u_iO_j\rangle$ is zero when $i \neq j$ so:

$$|\Omega\rangle = \sum_i d_i|u_iO_i\rangle \quad (142)$$

And we have:

$$p\{(u_i, O_i)\} = p\{u_i\} = p\{O_i\} = |c_i|^2 \quad (143)$$

We can write $|\Omega\rangle$ as:

$$|\Omega\rangle = \sum_i d_i|u_iO_i\rangle \neq |\Omega_1\rangle \otimes |\Omega_2\rangle \quad (144)$$

Where:

$$|d_i|^2 = |b_i|^2 = |c_i|^2 \quad (145)$$

So, the measurement is an entanglement between the system and the observer. And since the collapse of the state vector after the measurement is simply a matter of changing the probability distribution as we have seen, and after tossing the coin, the probability distributions for the three experiments: tossing the coin, the observer observing the result of the toss, and observing the observer observing the result will all change according to the result of the toss. So all the state vectors (138), (140), and (142) will collapse together and there is nothing mysterious about it: it is just a change in probability distributions!.

## 7. Time evolution of systems

The physical state of the system might change with time, so that means the state vector describing it might change with time, because the probability distributions of experiments might change with time. We will talk about time evolution of closed systems at first. First of all, what is the definition of a closed system? We will adopt the following definition:

A closed system is a system which satisfies that its characteristics are independent of time, meaning, when we study the system, it does not matter where we choose the origin of time, as long as we do not do a measurement on it.

Since the number of outcomes is the same in any moment of time we want to do the experiment, so at time $t$ we can represent the state vector of the experiment in the same vector space that we represented the state vector of it at time $t_0$.

We know that each observable is represented by a Hermitian operator, and the experiment we do to measure it has its events represented by orthonormal basis that is the eigenvectors of this operator. We will keep the bases representing all the experiments the same, and see how must the state vector change with time to keep satisfying that it is the state vector for the closed system. So, for the observable $\hat{A}$ that $\{|u_j\rangle\}_{j=1}^{N}$ are its eigenvectors we can keep them the same, but then the state vector might change in general. So the state vector at time $t$ is of the form:

$$|\Omega(t)\rangle = \sum_{j=1}^{N} c_j(t)|u_j\rangle \qquad (146)$$

Where $t$ is the time elapsed after the moment $t_0 = 0$.

We will search for an operator $\hat{U}$ such that, if we start the system in any initial state $|\Omega(0)\rangle$, then its state after a time $t$ is given by:

$$|\Omega(t)\rangle = \hat{U}(t)|\Omega(0)\rangle \qquad (147)$$

But since the state vector is a sample space vector, hence it is normalized, so:

$$\langle\Omega(t)|\Omega(t)\rangle = 1 \qquad (148)$$

So:

$$\langle\Omega(0)|\hat{U}^\dagger(t)\hat{U}(t)|\Omega(0)\rangle = 1 \qquad (149)$$

This can be done by choosing $\hat{U}$ to be isometric:

$$\hat{U}^\dagger(t)\hat{U}(t) = I \qquad (150)$$

Where $I$ is the identity operator. And as is known, this implies that this operator is linear[5].
Furthermore, from (147) we see that:

$$\hat{U}(0) = I \qquad (151)$$

It is a well known fact[1] that from the equations (150) and (151) in addition to (147) we can deduce that:

$$i\hbar \frac{d|\Omega\rangle}{dt} = \hat{H}|\Omega\rangle \qquad (152)$$

Where:

$$\hat{H}^\dagger = \hat{H} \qquad (153)$$

which means that $\hat{H}$ is a Hermitian operator.

Now, what about systems that are not closed?
We have to find an equation that describes the system in general, whether closed or not, and which becomes identical to (153) when the system is closed. For this, we can define an operator $\hat{H}$ for each system that satisfies:

1- $\hat{H}^\dagger = \hat{H}$

2- $i\hbar \frac{d|\Omega\rangle}{dt} = \hat{H}|\Omega\rangle$

3- When the system is closed, $\hat{H}$ is deduced from an isometric evolution with time of the state vector.

## 8. Treating a simple system as a composite system

If we have a system, and $E_1$ is an experiment of the class $C_N$ that we can do on it, and the state vector for the experiments of this class for this system is:

$$|\Omega_1\rangle = \sum_{i=1}^{N} c_i|u_i\rangle \in H_1 \qquad (154)$$

While $E_2$ is another experiment we can do on the system, and it is of the class $C_M$ where $N \neq M$. Where we assume that the state vector of this system for this class of experiments is:

$$|\Omega_2\rangle = \sum_{j=1}^{M} b_j|t_j\rangle \in H_2 \qquad (155)$$

Each experiment has its probability distribution, which we will assume it is independent from the other experiment.
As we saw, the sample space vector of the composite experiment $E$ which we can do on the system which is doing $E_1$ then doing $E_2$ on it is:

$$|\Omega\rangle = |\Omega_1\rangle \otimes |\Omega_2\rangle = \left(\sum_{i=1}^{N} c_i|u_i\rangle\right) \otimes \left(\sum_{j=1}^{M} b_j|t_j\rangle\right) \qquad (156)$$

And since it is a vector in a $N.M$ dimensional Hilbert space, then we can consider it as the state vector of the system for the class $C_{N.M}$ ensuring of course that we represent the experiments of this class by the bases that ensures that $|\Omega\rangle$ is a state vector.

But from the above we see that we can think of the system as equivalent to two separate non-interacting systems where the state vector of the first is given by (154), and the state vector of the second is given by (155). So the state vector of the composite system will be given by (156), since they are non-interacting.

# 9. Continuous probability distributions

What if we want the experiment to tell us about the position of some particle? Clearly, the way we used when talking about representing events by vectors is of no use here, because we are dealing with continuous probability distributions. So, we have to update our tools a little.

We will work first with a special example, then generalize. Let our example be a particle moving on a line.

Here, we will represent events by vectors in an infinite dimensional Hilbert space, because we have infinite values of $x$. And in this space, we will represent the observable $x$ by the operator $\hat{X}$, which satisfies:

$$\hat{X}|x\rangle = x|x\rangle \tag{157}$$

And we will demand the inner product in this space to satisfy:

$$\langle x|x'\rangle = \delta(x-x') \tag{158}$$

Furthermore, we demand this inner product to satisfy also:

For any $|\psi\rangle$ and any well behaved complex function $f(x)$:

$$\langle \psi | \left( \int_a^b dx f(x) |x\rangle \right) = \int_a^b dx f(x) \langle \psi | x \rangle \tag{159}$$

$$\left( \int_a^b dx f(x) \langle x | \right) |\psi\rangle = \int_a^b dx f(x) \langle x | \psi \rangle \tag{160}$$

Where: $a,b \in R \cup \{-\infty,+\infty\}$.

We can write any $|\psi\rangle$ as:

$$|\psi\rangle = \int_{-\infty}^{+\infty} dx\, c(x) |x\rangle \tag{161}$$

Because then:

$$\langle x'|\psi\rangle = \int_{-\infty}^{+\infty} dx\, c(x) \langle x'|x\rangle$$
$$= \int_{-\infty}^{+\infty} dx\, c(x) \delta(x-x') = c(x') \tag{162}$$

So:

$$|\psi\rangle = \int_{-\infty}^{+\infty} dx |x\rangle\langle x|\psi\rangle \tag{163}$$

We can define $\int_a^b dx |x\rangle\langle x|$ where: $a,b \in R \cup \{-\infty,+\infty\}$, as follows:

For any $|\psi\rangle$:

$$\left( \int_a^b dx |x\rangle\langle x| \right) |\psi\rangle = \int_a^b dx |x\rangle\langle x|\psi\rangle \tag{164}$$

$$\langle \varphi | \left( \int_a^b dx |x\rangle\langle x| \right) = \int_a^b dx \langle \varphi | x\rangle\langle x| \tag{165}$$

So: $\forall |\psi\rangle$:

$$\left( \int_{-\infty}^{+\infty} dx |x\rangle\langle x| \right) |\psi\rangle = \int_{-\infty}^{+\infty} dx |x\rangle\langle x|\psi\rangle = |\psi\rangle = I|\psi\rangle \Rightarrow$$

$$\int_{-\infty}^{+\infty} dx |x\rangle\langle x| = I \tag{166}$$

So we have:

$$\langle \varphi|\psi\rangle = \langle \varphi|I|\psi\rangle = \langle \varphi| \int_{-\infty}^{+\infty} dx |x\rangle\langle x|\psi\rangle$$

$$= \int_{-\infty}^{+\infty} dx \langle \varphi|x\rangle\langle x|\psi\rangle \tag{167}$$

With all of that being said, now we can represent events as the following:

We represent the event:

$$A = \{x : x \in (a,b) \cup ... \cup (c,d)\} \tag{168}$$

(where $a,b,...,c,d \in R \cup \{-\infty,+\infty\}$ and the intervals $(i,j)$ might be open, closed or half open/half closed and we have a finite number of them, let us call it $N$) by a vector $|A\rangle$ in some infinite dimensional Hilbert space $H$ as follows:

1- $\quad p(A) = \langle A|A\rangle$

2- $\quad x \notin A \Rightarrow \langle x|A\rangle = 0$

We know that:

$$|A\rangle = I|A\rangle = \int_{-\infty}^{+\infty} dx |x\rangle\langle x|A\rangle$$

Thus:

$$|A\rangle = \int_{-\infty}^{a} dx |x\rangle\langle x|A\rangle + \int_{a}^{b} dx |x\rangle\langle x|A\rangle + ... +$$
$$+ \int_{c}^{d} dx |x\rangle\langle x|A\rangle + \int_{d}^{+\infty} dx |x\rangle\langle x|A\rangle \tag{169}$$

But all integrals except the ones in which are over intervals spanned by the event $A$ are equal to zero because in the intervals that do not satisfy that, we have $\langle x|A\rangle = 0$ so:

$$|A\rangle = \int_{a}^{b} dx |x\rangle\langle x|A\rangle + ... + \int_{c}^{d} dx |x\rangle\langle x|A\rangle \tag{170}$$

Where we are integrating only over intervals spanned by $A$

and this is the meaning we will give to the former equation throughout this paper. We can see that:

$$|\Omega\rangle = \int_{-\infty}^{+\infty} dx |x\rangle\langle x|\Omega\rangle \quad (171)$$

Since:

$$A' = \Omega \setminus A \quad (172)$$

That yields:

$$|\Omega\rangle = |A\rangle + |A'\rangle \quad (173)$$

Now:

$$p(A) = \langle A|A\rangle = \int_{-\infty}^{+\infty} dx \langle A|x\rangle\langle x|A\rangle$$
$$= \int_a^b dx \langle A|x\rangle\langle x|A\rangle + ... +$$
$$+ \int_c^d dx \langle A|x\rangle\langle x|A\rangle \quad (174)$$

But from (173) we know that:

$$|\Omega\rangle = |A\rangle + |A'\rangle \Rightarrow$$
$$\langle x|\Omega\rangle = \langle x|A\rangle + \langle x|A'\rangle \quad (175)$$

We know that when $x \in A$ we have: $\langle x|A'\rangle = 0$ so substituting in (175) gives:

$$x \in A \Rightarrow \langle x|\Omega\rangle = \langle x|A\rangle \quad (176)$$

Using the former equation in (174) we get:

$$p(A) = \int_a^b dx \langle \Omega|x\rangle\langle x|\Omega\rangle + ... + \int_c^d dx \langle \Omega|x\rangle\langle x|\Omega\rangle$$
$$= \int_a^b dx |\Omega(x)|^2 + ... + \int_c^d dx |\Omega(x)|^2 \quad (177)$$

Where:

$$\Omega(x) = \langle x|\Omega\rangle \quad (178)$$

So the probability of $x$ being in the interval $(a,b) \cup ... \cup (c,d)$ is:
$\int_a^b dx |\Omega(x)|^2 + ... + \int_c^d dx |\Omega(x)|^2$ so $|\Omega(x)|^2$ is the probability density.

We can generalize this process to any continuous observable. And if we want to measure another incompatible observable, we do as we did in the case of discrete variables, meaning we represent it by another basis.

It is fairly easy to generalize for three dimensions.

Also here, we will represent events by vectors in an infinite dimensional Hilbert space. And we will suppose that there is in this space the vectors $\{|\vec{r}\rangle\}$ which satisfy:

$$\langle \vec{r}|\vec{r}'\rangle = \delta(\vec{r} - \vec{r}') \quad (179)$$

Where:

$$\iiint f(\vec{r}) \delta(\vec{r} - \vec{r}') d\tau = f(\vec{r}') \quad (180)$$

Where the integration is evaluated over the whole of space and $\vec{r}$ is the position vector of the particle, and $d\tau$ is the infinitesimal volume around $\vec{r}$. Furthermore we demand the inner product in this space to satisfy:

For any $|\psi\rangle$ and any well behaved complex function $f(\vec{r})$:

$$\left\langle \psi \middle| \iiint_\tau d\tau f(\vec{r}) |\vec{r}\rangle \right\rangle = \iiint_\tau d\tau f(\vec{r}) \langle \psi|\vec{r}\rangle \quad (181)$$

$$\left( \iiint_\tau d\tau f(\vec{r}) \langle \vec{r}| \right) |\psi\rangle = \iiint_\tau d\tau f(\vec{r}) \langle \vec{r}|\psi\rangle \quad (182)$$

Where $\tau$ may be all of space or only some part of it.

And we define $\iiint_\tau d\tau |\vec{r}\rangle\langle \vec{r}|$ where $\tau$ may be the whole of space or only some part of it, as follows: for any $|\psi\rangle$:

$$\left( \iiint_\tau d\tau |\vec{r}\rangle\langle \vec{r}| \right) |\psi\rangle = \iiint_\tau d\tau |\vec{r}\rangle\langle \vec{r}|\psi\rangle \quad (183)$$

$$\left\langle \psi \middle| \left( \iiint_\tau d\tau |\vec{r}\rangle\langle \vec{r}| \right) \right. = \iiint_\tau d\tau \langle \psi|\vec{r}\rangle\langle \vec{r}| \quad (184)$$

From all of the above we can prove, in a similar manner to what we did in the one dimensional case, that:

For any $|\psi\rangle$ and $|\varphi\rangle$ we have:

$$|\psi\rangle = \iiint d\tau |\vec{r}\rangle\langle \vec{r}|\psi\rangle \quad (185)$$

$$\langle \varphi|\psi\rangle = \iiint d\tau \langle \varphi|\vec{r}\rangle\langle \vec{r}|\psi\rangle \quad (186)$$

$$\iiint d\tau |\vec{r}\rangle\langle \vec{r}| = I \quad (187)$$

Where the integration extends over all of space, and $I$ is the identity operator.

With these tools, we can continue exactly as we did in the one dimensional case.

## 10. Quantum mechanics

Now with these concepts at hand, we need no assumptions in quantum mechanics, just we need to apply them. We will take as an example the component of the spin of an electron along the z-axis[1].

If we created an electron somehow, in general, when we turn on a magnetic field in the z-direction, we might find the component of the spin of the electron either up or down with a certain probability depending on the initial state. If the electron was originally up or originally down, then we find it after turning on the magnetic field certainly up/down as is known [1]. But there are states that we sometimes get up and other times get down so they are different than the electron being up or being down before we turn on the magnetic field[1]. So how do we explain that according to the new formulation of probability theory? In this new interpretation of quantum mechanics, we see that quantum mechanics is just a statistical theory in the same way classical experiments are. To understand the experiment of measuring the component of the spin of an electron, we will compare it to a classical experiment which is measuring the Headness or the Tailsness of a coin. What we mean by the statement: the result of the experiment is Heads/Tails, is that after we toss the coin, the upper surface of the coin after it stabilizes on a horizontal surface is Heads/Tails. So, before tossing the coin, i.e. when it is still in my hand, there is no meaning of the question "is the coin Heads or Tails?". All we can talk about is the sample space of the experiment. And in this case we saw that the state vector of the coin (which represents the sample space) is of the form:

$$|\Omega\rangle = \alpha_H |H\rangle + \alpha_T |T\rangle : |\alpha_H|^2 + |\alpha_T|^2 = 1 \quad (188)$$

Now, after tossing the coin, we have some definite result, either Heads or Tails. And as we saw before, we can represent the sample space vector of any experiment of the same class that we can do on the coin after tossing, i.e. the state vector after the measurement, by (if the result was say Heads):

$$|\Omega'\rangle = |H\rangle \quad (189)$$

Which means that if we read the result of tossing, we will find it definitely Heads.

In the same way, before turning on the magnetic field, there is no meaning of the question "is the component of the spin of the electron up or down?" but the state vector before the measurement (which represents the sample space of measuring the component) is of the form:

$$|\psi\rangle = \alpha_u |u\rangle + \alpha_d |d\rangle : |\alpha_u|^2 + |\alpha_d|^2 = 1 \quad (190)$$

But after turning on the magnetic field (the equivalent of tossing the coin in the former experiment), we will have a definite result, let us say up, so the state vector after the measurement can be written as:

$$|\psi'\rangle = |u\rangle \quad (191)$$

And that represents the sample space for any experiment of the same class after the measurement.

## 11. What about Bell's theorem?

In fact this new interpretation is in total agreement with Bell's theorem.

As we saw when we spoke of incompatible observables, they are in this interpretation, observables that cannot be measured in the same experiment. And that any two observables that their commutator is not zero are incompatible.

Furthermore, we saw that if the sample space of the experiment is not already an eigenvector of the observable we want to measure, then there is no meaning of the question "what was this property before doing the experiment?". So, when we have two electrons in the singlet state, and since according to this interpretation, the state vector in the singlet state is nothing more than a representation of the sample space of the experiment of measuring the spins of the electrons, and this sample space is represented by:

$$|\psi\rangle = \frac{1}{\sqrt{2}}(|ud\rangle - |du\rangle) \quad (192)$$

It means that we cannot say before measuring the spins that the spins were opposite because it is like saying the coin is Heads while it is still in my hand.

Also, the three components of the spin of an electron are represented by incompatible observables, so they cannot be measured together. And since there is no meaning of the question: 'what is the component before measuring it?", so we cannot say that the three components of the spin of the electron can have values together.

From all of the above we see that Bell's theorem is actually in favour of this interpretation and supports it.

## FINAL THOUGHTS

In the end, we see that this interpretation of quantum mechanics follows naturally from the mathematics that I have described, and that is why I call this interpretation "the natural interpretation of quantum mechanics", and I have pub-lished it for the first time in the International Journal of Theoretical and Mathematical Physics[6].

## ACKNOWLEDGEMENT

I wish to acknowledge Diaa Fadel and Ammar Atrash for their help in typing this document.